\documentclass[twocolumn]{aastex631}
\usepackage{amsmath,amsthm,mathtools}
\usepackage{tabularray}
\usepackage{comment}

\received{December 22, 2023}
\revised{May 13, 2024}
\accepted{May 13, 2024}

\submitjournal{ApJ}

\shortauthors{Thang et al.}

\begin{document}

\title{\textbf{\LARGE Evidence of Grain Alignment by Magnetically Enhanced Radiative Torques from Multiwavelength Dust Polarization Modeling of HL Tau}}
\correspondingauthor{Nguyen Tat Thang}
\email{thangnguyentat3@gmail.com}

\author[0009-0004-1141-0678]{Nguyen Tat Thang}
\affil{Department of Astrophysics, Vietnam National Space Center, Vietnam Academy of Science and Technology, 18 Hoang Quoc Viet, Hanoi, Vietnam}
\affil{Simons Astrophysics Group at IFIRSE (SAGI)/Institute For Interdisciplinary Research in Science and Education (IFIRSE), ICISE, 07 Science Avenue, Ghenh Rang Ward, 55121 Quy Nhon City, Binh Dinh Province, Vietnam}

\author[0000-0002-2808-0888]{Pham Ngoc Diep}
\affiliation{Department of Astrophysics, Vietnam National Space Center, Vietnam Academy of Science and Technology, 18 Hoang Quoc Viet, Hanoi, Vietnam}
\affiliation{Graduate University of Science and Technology, Vietnam Academy of Science and Technology, 18 Hoang Quoc Viet, Hanoi, Vietnam}

\author[0000-0003-2017-0982]{Thiem Hoang}
\affiliation{Korea Astronomy and Space Science Institute, Daejeon 34055, Republic of Korea}
\affiliation{Department of Astronomy and Space Science, University of Science and Technology, 217 Gajeong-ro, Yuseong-gu, Daejeon, 34113, Republic of Korea}

\author[0000-0002-6488-8227]{Le Ngoc Tram}
\affiliation{Max-Planck-Institut für Radioastronomie, Auf dem Hügel 69, 53121, Bonn, Germany}

\author[0000-0002-5913-5554]{Nguyen Bich Ngoc}
\affiliation{Department of Astrophysics, Vietnam National Space Center, Vietnam Academy of Science and Technology, 18 Hoang Quoc Viet, Hanoi, Vietnam}
\affiliation{Graduate University of Science and Technology, Vietnam Academy of Science and Technology, 18 Hoang Quoc Viet, Hanoi, Vietnam}

\author[0000-0002-4372-5509]{Nguyen Thi Phuong}
\affiliation{Department of Astrophysics, Vietnam National Space Center, Vietnam Academy of Science and Technology, 18 Hoang Quoc Viet, Hanoi, Vietnam}
\affil{Simons Astrophysics Group at IFIRSE (SAGI)/Institute For Interdisciplinary Research in Science and Education (IFIRSE), ICISE, 07 Science Avenue, Ghenh Rang Ward, 55121 Quy Nhon City, Binh Dinh Province, Vietnam}

\author[0000-0001-9654-8051]{Bao Truong}
\affiliation{Korea Astronomy and Space Science Institute, Daejeon 34055, Republic of Korea}
\affiliation{Korea University of Science and Technology, 217 Gajeong-ro, Yuseong-gu, Daejeon, 34113, Republic of Korea}

\begin{abstract}
Atacama Large Millimeter/Submillimeter Array has revolutionized the field of dust polarization in protoplanetary disks across multiple wavelengths. Previous observations and empirical modeling suggested multiple mechanisms of dust polarization toward HL Tau, including grain alignment and dust scattering. However, a detailed modeling of dust polarization based on grain alignment physics is not yet available. Here, using our updated POLARIS code, we perform numerical modeling of dust polarization arising from both grain alignment by Magnetically Enhanced Radiative Torque mechanism and self-scattering to reproduce the HL Tau polarization observed at three wavelengths 0.87, 1.3, and 3.1$\,$mm. Our modeling results show that the observed multi-wavelength polarization could be reproduced only when large grains contain embedded iron inclusions and those with slow internal relaxation must have wrong internal alignment (i.e., the grain's major axis parallel to its angular momentum). The abundance of iron embedded inside grains in the form of clusters is constrained to be $\gtrsim 16$\%, and the number of iron atoms per cluster is $N_{\rm cl} \sim 9\times10^2$. Maximum grain sizes probed at wavelengths $\lambda$ = 0.87, 1.3, and 3.1$\,$mm are constrained at $\sim$ 60, 80, and 90$\,\mu$m, respectively.
\end{abstract}


\section{Introduction} 

Magnetic fields are thought to play a crucial role in protoplanetary disks in driving magnetorotational instability \citep{1959JETP....9..995V, 1991ApJ...376..214B}, disk formation (e.g., magnetic braking and angular momentum transport) \citep{1991ApJ...376..214B, 1998RvMP...70....1B, 1994ASPC...54...73H}, protostellar outflow launching \citep{1982MNRAS.199..883B, 2021AAS...23832901S}, and planet formation \citep{1998ApJ...508..707Q, 2003A&A...411..623F, 2009IAUS..259..249J}. Polarized thermal emission from aligned grains has been known as the leading tool to investigate the magnetic fields in the interstellar medium (ISM) and molecular clouds (MCs). Therefore, dust polarization is expected to be a powerful tool in studying magnetic fields \citep{2014Natur.514..597S} and dust properties \citep{2015ApJ...809...78K} in disks. Since the emergence of the Atacama Large Millimeter/submillimeter Array (ALMA), the field of (sub)millimeter-wavelength disk polarization has progressed significantly with the polarization signatures being observed to incredible detail and accuracy \citep{2017ApJ...851...55S, 2017ApJ...844L...5K, 2018ApJ...855...92C, 2019ApJ...877L...2H, 2019ApJS..245....2S, 2023ApJ...947L...5T}.

HL Tau is a perfect example to illustrate the evolutionary process of the field. It is a young protoplanetary disk (SED class I/II, age $\sim\,$1 Myr) \citep{2004ApJ...616..998W} located in the Taurus molecular cloud, which is $\sim\,140\,$pc from Earth \citep{2004AJ....127.1029R}. It is one of the earliest disks observed with polarimetry by Combined Array for Research in Millimeter-wave Astronomy (CARMA) and Submillimeter Array (SMA). Its polarized emission was used to study the disk's magnetic fields, supposing grains are magnetically aligned \citep{2014Natur.514..597S}. However, the study concluded that a simple magnetic field structure could not fully explain the detected polarization. Coming to the age of ALMA observations, the disk has been revealed to have a transition of polarization morphology from azimuthal to parallel to the disk's minor axis when probing at longer to shorter wavelengths \citep{2017ApJ...844L...5K, 2017ApJ...851...55S}. These new observations essentially raise questions about the validity of tracing magnetic fields by dust polarization within the disk.

The complex polarization patterns indicate the possibility of multiple mechanisms involved in producing the dust-polarized emission of HL Tau at (sub)millimeter wavelengths. Indeed, previous studies have suggested that a superposition of self-scattering and thermal emission from azimuthally-aligned prolate grains can explain such observed polarized emission \citep{2019MNRAS.483.2371Y, 2021ApJ...908..153M, 2022MNRAS.512.3922L}. Self-scattering is caused by dust grains with sizes comparable to the observed wavelengths scattering and polarizing incident lights from the thermal emission of other grains \citep{2015ApJ...809...78K}. Grain alignment, on the other hand, has elongated grains being systematically aligned with a preferred direction and emitting polarized thermal electromagnetic waves. Nevertheless, which mechanisms exactly induce grain alignment in such a way is still a mystery since most of the proposed mechanisms (i.e., magnetic alignment, radiative alignment, aerodynamic alignment) fail to interpret the observed disk polarization \citep{2017ApJ...839...56T, 2019MNRAS.483.2371Y, 2021ApJ...908..153M}.

Grain alignment in the ISM and MCs has been studied with great success and made dust polarization an invaluable tool to trace magnetic fields within the environment \citep{2015ARA&A..53..501A, 2019FrASS...6...15P}. However, its applicability to the protostellar environment is complex due to the presence of very large grains (grain size $a\,\gtrsim\,\text{10}\,\mu$m) and significant gas randomization effect. Recently, \cite{2022AJ....164..248H} considered a more realistic grain alignment model within such an environment in contrast to previous modelings where only perfect alignment is included. The process of grain alignment therein consists of the alignment of the axis of maximum moment of inertia ($\boldsymbol{\hat{a_1}}$) with the angular momentum ($\boldsymbol{J}$), so-called internal alignment (IA); and the alignment of $\boldsymbol{J}$ with a preferred direction, so-called external alignment (EA). Internal alignment can be achieved by internal relaxation (INR), e.g., Barnett relaxation \citep{1993ApJ...418..287R} and inelastic relaxation \citep{2003A&A...398..809M}. External alignment, on the other hand, can be accomplished by grain precession (e.g., Larmor precession, radiative precession, and mechanical precession). Grains achieving external alignment with Larmor precession, radiative precession, or mechanical precession have $\boldsymbol{J}$ aligned along the direction of the magnetic field ($\boldsymbol{B}$), radiation beam ($\boldsymbol{k}$), or gas flow ($\boldsymbol{\nu}$). The mechanisms behind such alignments are called magnetic alignment (\citealp{2022ApJ...928..102H}), radiative alignment ($k$-RAT; \citealp{2007MNRAS.378..910L}), or mechanical alignment ($v$-MET; \citealp{2007ApJ...669L..77L}), respectively. 

Recently, \cite{2023MNRAS.520.3788G} incorporated such grain alignment model (i.e., realistic internal alignment and external alignment model) into POLARIS \citep{2016A&A...593A..87R} taking into account only the grain alignment mechanism of Magnetically-enhanced Radiative Torque (MRAT) \citep{2016ApJ...831..159H} which was specifically applied to study thermal dust polarization towards protostellar cores. With this mechanism, grains can be aligned with the disk magnetic fields by radiative torques (RATs; \citealp{1976Ap&SS..43..291D, 1997ApJ...480..633D,2008MNRAS.388..117H}) with enhanced efficiencies by (super)paramagnetic relaxation \citep{1951ApJ...114..206D, 2008MNRAS.388..117H, 2016ApJ...831..159H}. 

In this paper, our main goal is to test the validity of grain alignment with the MRAT mechanism and put a strong constraint on grain magnetic properties as well as grain growth in HL Tau. To do so, we perform multi-wavelength modeling of disk polarization using the updated POLARIS \citep{2023MNRAS.520.3788G} for three well-studied ALMA Bands 3, 6, and 7 corresponding to the wavelengths $\lambda$ = 3.1, 1.3, and 0.87$\,$mm.

The structure of this paper is as follows. In Section \ref{model}, we describe the disk and dust model used in our modeling. Subsequently, our main results including synthetic polarization from thermal aligned dust emission with MRAT mechanism; and from the mixture of grain alignment and self-scattering are shown in Section \ref{ncl_pol_result} and \ref{resultf}, respectively. We further discuss the implications, grain alignment properties, and grain growth in Section \ref{sec:discussions}. Finally, the main findings are summarized in Section \ref{summary}.

\begin{figure*}
    \centering
    \includegraphics[scale=0.5]{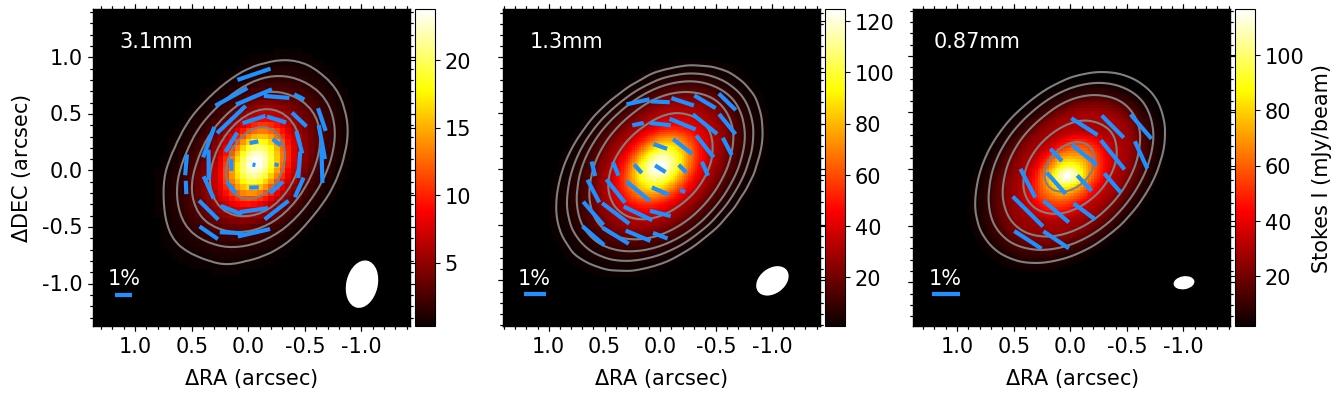}
    \caption{Left to right: Stokes $I$ intensity overlaid with ALMA polarimetric observations of HL Tau at Band 3, Band 6, and Band 7. The line segments represent the polarization vectors and are set to follow the scale of polarization degree. The ones with polarized intensity below signal-to-noise of 3 are not plotted. A reference vector of 1$\%$ is shown in the lower-left corner of each panel. The contours represent the intensity levels of (10, 25, 50, 100, 200) $\times$ the noise level, which is $\sigma_{\rm I}$ = 58, 170, and 409$\,\mu$Jy/beam at $\lambda$ = 3.1, 1.3, and 0.87$\,$mm, respectively. The beam sizes and their position angles are ($0.45\arcsec \times 0.29\arcsec$, $-12.8^\circ$), ($0.33\arcsec \times 0.24\arcsec$, $-53.4^\circ$), and ($0.2\arcsec \times 0.13\arcsec$, $-80.9^\circ$), respectively.}
    \label{observation}
\end{figure*}

\begin{table*}
    \centering
    \caption{Information of the observational data used in this paper}
    \label{table::obs}
    \begin{tblr}{c c c c c c c c c c}
        \hline
            \textbf{Band} & \textbf{$\lambda$} & \textbf{$\sigma_{I}$} & \textbf{$\sigma_{PI}$} & \textbf{BMAJ} & \textbf{BMIN} & \textbf{BPA} & \textbf{Proj.code} & \textbf{P.I.} & \textbf{Reference} \\ 
             & (mm) & $\left(\frac{\mu \rm Jy}{\rm beam}\right)$ & $\left(\frac{\mu \rm Jy}{\rm beam}\right)$ & (arcsec) & (arcsec) & (deg) & & & \\
             (1) & (2) & (3) & (4) & (5) & (6) & (7) & (8) & (9) & (10)\\
        \hline
            3 & 3.1 & 58 & 22 & 0.45 & 0.29 & $-12.8$ & 2016.1.00115.S & Kataoka, Akimasa & (I), (II), (III) \\
            6 & 1.3 & 170 & 37 & 0.33 & 0.24 & $-53.4$ & 2016.1.00162.S & Stephens, Ian & (II), (III) \\
            7 & 0.87 & 409 & 92 & 0.20 & 0.13 & $-80.9$ & 2019.1.01051.S & Stephens, Ian & (III) \\
        \hline    
    \end{tblr}

    \begin{minipage}{18cm}
    \vspace{0.3cm}
    \small  \textbf{Note} (1) Name of the observation band; (2) Central wavelength; (3) Noise level of Stokes $I$ emission; (4) Noise level of polarized emission; (5) Beam's major axis; (6) Beam's minor axis; (7) Beam's position angle; (8) Project code; (9) Principal Investigator; (10) Associated reference. \newline
    \textbf{References.} (I) \citet{2017ApJ...844L...5K}; (II) \citet{2017ApJ...851...55S}; (III) \citet{2024MNRAS.528..843L}.
    \end{minipage}
    \vspace{0.5cm}
\end{table*}

\section{Dust and Disk Models}
\label{model}
The outline of our model is briefly described as follows. Firstly, we construct a dust model and a best-fitted disk model to the Stokes $I$ emissions obtained by ALMA at the three wavelengths $\lambda$ = 3.1, 1.3, and 0.87$\,$mm. Then, by separately measuring direct thermal emission from grains aligned by MRAT and scattered emission from self-scattering\footnote{The updated POLARIS code described in \citet{2023MNRAS.520.3788G} was built up from a rather old version of POLARIS, whose dust scattering feature is less well-tested. Hence, for the Monte-Carlo radiative transfer calculation of dust self-scattering, we use the POLARIS code taken from  \url{https://github.com/polaris-MCRT/POLARIS/tree/master-old} instead.}, we combine their Stokes parameters to obtain the combined linear polarization from the dust emission. 

Figure \ref{observation} shows ALMA Stokes $I$ intensity overlaid with polarization vectors of HL Tau disk at $\lambda$ = 3.1, 1.3, and 0.87$\,$mm. The details of the information regarding the observational data used in this paper are summarized in Table \ref{table::obs}. The observational data is taken directly from the ALMA archive, which has been reduced and calibrated by ALMA staff. We only further perform de-biasing the polarized intensity and polarization degree via the method described in \citet{2014ApJS..213...13H} and \citet{2015JAI.....450005H}. The ALMA minimum detectable polarization degree is expected at 0.1\%, and defined as three times the systematic calibration uncertainty. Hence, we use the error of 0.033\% when the polarization degree uncertainty is less than this value. 

The notable features of the observations are (1) the transition of the polarization patterns from azimuthal to parallel to the disk's minor axis from 3.1 to $0.87\,\rm mm$ and (2) the polarization degree is generally at the level of $\sim$1\% for the whole three wavelengths.

\subsection{Dust Model}
We adopt a dust model with compositions from \cite{2018ApJ...869L..45B}, where dust grains are a mixture of astronomical silicate, troilite, refractory organics, and water ice. The refractive index used in the calculation is from \cite{2003ARA&A..41..241D} for astronomical silicate, \cite{1996A&A...311..291H} for troilite, and \cite{2008JGRD..11314220W} for refractory organics and water ice. Assuming compact grains, the mean refractive index of the mixture is then calculated with the effective medium theory using Bruggeman rule \citep{1998asls.book.....B}. 

The grain sizes are assumed to follow a power-law distribution with a power index $q = -3.5$ \citep{1977ApJ...217..425M}, a maximum grain size $a_{\rm max}$, and a fixed minimum grain size $a_{\rm min}$ = 0.05 $\mu$m. We assume the grain population to be the same throughout the disk. As discussed in \cite{2022MNRAS.512.3922L}, the value of $a_{\rm max}$ does not converge for the three ALMA wavelengths (i.e., there is no single value could satisfy the polarization constraint at the three wavelengths at once). Therefore, we leave $a_{\rm max}$ as a wavelength-dependent parameter. We assume the Rayleigh regime \citep{1985ApJ...290..211L} for the polarization calculation. This approximation holds when the size parameter $x\,=\,2\pi a/\lambda\,<<\,1$. Indeed, later on, this is shown to be the case with the maximum grain sizes found to be $\sim$60, 80, and 90$\,\mu$m for the three probing wavelengths 0.87, 1.3, and 3.1$\,$mm, respectively.

\begin{table*}
    \centering
    \caption{Key Parameters of the Dust Model}
    \label{table::dust}
    \begin{tblr}{p{8cm} c c}
        \hline
            \textbf{Parameter} & \textbf{Denotation} & \textbf{Value} \\ 
        \hline
            Dust mixture & - & silicate, organics, water ice, troilite \\
            \hline
            Grain size distribution & dn/da & $Ca^{-3.5}$ \\
            Minimum grain size & $a_{\rm min}$ & 0.05$\, \mu$m \\
            Maximum grain size & $a_{\rm max}$ & $\lambda$ - dependent \\
            \hline
            Grain aspect ratio & $s$ & 1/2 (ray-tracing) or 1 (MCRT) \\
            \hline
            Iron fraction of PM grains & $f_{\rm p}$ & 0.1 \\
            \hline
            Iron cluster volume filling factor of SPM grains & $\phi_{\rm sp}$ & 0.1 \\
            Number of iron atoms per cluster of SPM grains & $N_{\rm cl}$ & [ 20, $10^5$ ] \\
            \hline
            IA efficiency at low-$J$ and slow INR & $Q_{\rm X,low-J}$ & $-0.4$ (wrong IA) or 0.4 (right IA) \\
            IA efficiency at high-$J$ and slow INR & $Q_{\rm X,high-J}$ & 0.4 \\
        \hline    
    \end{tblr}

    \begin{minipage}{16cm}
    \vspace{0.3cm}
    \small  \textbf{Notes.} PM/SPM grains denote the grains made of paramagnetic/superparamagnetic materials. IA denotes internal alignment process, and INR denotes internal relaxation.
    \end{minipage}
    \vspace{0.5cm}
\end{table*}

We assume two different grain shapes for direct thermal dust emission ray tracing and Monte-Carlo radiative transfer (MCRT) calculation (i.e., radiation field, radiative heating, and dust scattering calculations). For dust emission ray tracing, which is used to trace aligned grain emission, oblate grains with aspect ratio $s$ ($0<s<1$) are assumed. Note that grain shape has a strong effect on the degree of polarization. Grains with smaller $s$ are expected to produce higher polarization degrees. However, since the exact grain geometry in the disk environment is not yet constrained, we choose an average value of $s=1/2$ for simplicity. The grains' optical properties including extinction and RAT efficiencies are then pre-calculated with DDSCAT \citep{1994JOSAA..11.1491D}. For the MCRT calculation, we adopt spherical grains (i.e., $s\,=\,$1), whose dust optical properties can be analytically calculated with Mie theory \citep{2012mith.book...53W}. Though the two-grain shapes are not consistent, we neglect the effect of scattering with non-spherical grains for simplicity. As pointed out by \cite{2020A&A...638A.116K}, assuming perfect compact spherical grains yields quantitative errors compared to non-spherical grains. However, the errors are negligible in the Rayleigh regime (i.e., spherical grains are still an acceptable representation).

Self-scattering-induced polarization is achieved by treating dust as photon emitters in dust scattering radiative transfer simulation. The controlling parameter of scattering is the maximum grain size, $a_{\rm max}$. Self-scattering polarization is most effective when $a_{\rm max}\,\sim\,\lambda/2\pi$ \citep{2015ApJ...809...78K}.

In the present work, grain alignment is considered to be induced only by MRAT. Under the framework of MRAT, grains can efficiently be aligned with magnetic fields owing to the enhanced efficiency by RATs and (super)paramagnetic relaxation due to iron inclusion within the grains \citep{2008ApJ...676L..25L, 2016ApJ...831..159H}. Recent observations show more evidence for iron inclusion in the form of clusters in different environments such as in protostellar core \citep{2023MNRAS.520.3788G}, filament \citep{2023ApJ...953...66N}, and the envelope of evolved stars \citep{2023arXiv230801215T}. RATs can spin grains up to suprathermal rotation (so-called high-$J$ attractors) or de-spin them down to thermal rotation (so-called low-$J$ attractors). Grains with iron inclusion are dubbed paramagnetic material (PM) or superparamagnetic material (SPM). Considering paramagnetic grains, we assign the value of the fraction of iron atoms diffusely distributed within grains as $f_{\rm p} = 0.1$. For superparamagnetic grains, we fix the volume filling factor of iron clusters to $\phi_{\rm sp}$ = 0.1, which corresponds to an iron abundance of $\sim$30\% presented in the form of iron clusters\footnote{In reality, $\phi_{\rm sp}$ can be as large as 0.3, which corresponds to an iron abundance of $\sim$100\% presented in the form of iron clusters \citep{2016ApJ...831..159H}} and vary the number of iron atoms per cluster $N_{\rm cl}$ to describe different degree of grain alignment enhancement due to superparamagnetic relaxation. The possible value of $N_{\rm cl}$ spans from $\sim$ 20 to $10^5$ \citep{1967ApJ...147..943J}.

Note that for internal relaxation of grains, although the case of RATs quickly bringing $\boldsymbol{J}$ to align with $\boldsymbol{\hat{a_1}}$ (fast internal relaxation, \citealp{2021ApJ...908...12L}) has been well-studied, the case of internal relaxation timescale being much longer than gas damping timescale (slow internal relaxation) is not as well-defined \citep{2022AJ....164..248H}. Hence, in this study, for the case of slow internal relaxation, we fix the values of internal alignment efficiency $Q_{\rm X}$ as follows. When grains are aligned at high-$J$ attractors, the internal alignment efficiency is $Q_{\rm X,high-J}$ = 0.4. On the other hand, when grains are at low-$J$ attractors, they can be aligned at either right or wrong internal alignment \citep{2009ApJ...697.1316H}. In the case of right internal alignment, $\boldsymbol{J}$ is aligned parallel to $\boldsymbol{\hat{a_1}}$, which coincides with the minor axis of the ellipsoidal grains. In the case of wrong internal alignment, their alignment direction is perpendicular instead. When $Q_{\rm X,low-J} = -0.5$, grains are defined to achieve perfect wrong internal alignment \citep{2023MNRAS.520.3788G}. We set $Q_{\rm X,low-J}$ = 0.4 for the case of grains aligning at right internal alignment, and $Q_{\rm X,low-J} = -0.4$ at wrong internal alignment. The choice of the values of $Q_{\rm X,high-J}$ and $Q_{\rm X,low-J}$ at wrong internal alignment is taken as the median value of the constraint which will be given in Section \ref{sINR}. The value of $Q_{\rm X,low-J}$ at right internal alignment is set to be similar to $Q_{\rm X,high-J}$ based on the assumption that $\boldsymbol{J}$ would have a negligible effect on the grains' internal alignment efficiency when they are aligned with slow internal relaxation. Throughout the paper, we will call wrong internal alignment/right internal alignment for short in referring to the case of grains aligned with wrong or right internal alignment when they are at low-$J$ attractors and aligned with slow internal relaxation. The procedure to calculate the alignment degree of dust grains with magnetic fields in POLARIS is nicely summarized in Figure 2 of \citet{2023MNRAS.520.3788G}.

A summary of the critical parameters used in our dust model is given in Table \ref{table::dust}. 

\begin{figure*}
    \centering
    \includegraphics[scale=0.5]{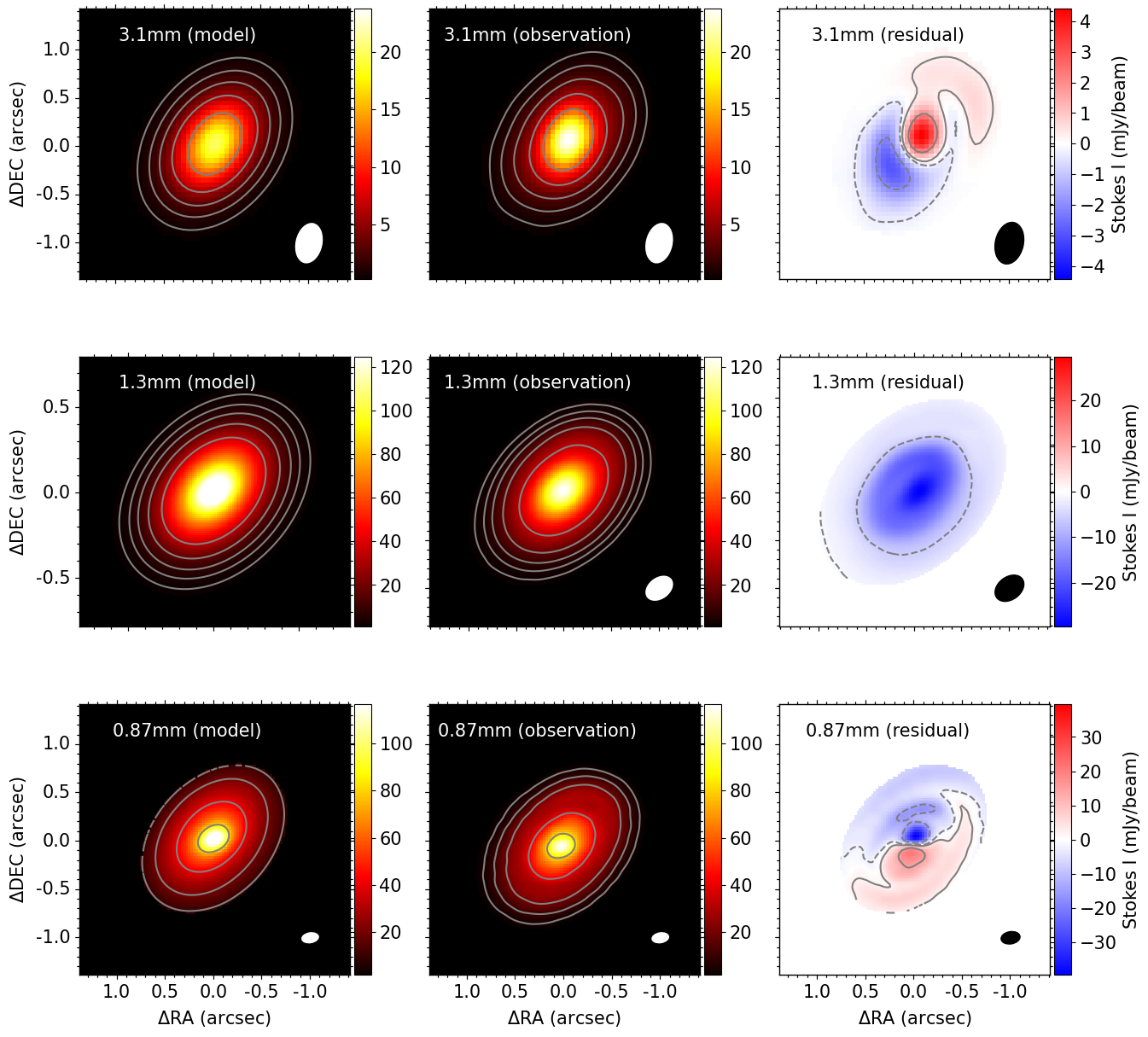}
    \caption{From left to right: Stokes $I$ emissions of model, observation, and residual (observation$-$model) at 3.1$\,$mm (top), 1.3$\,$mm (middle), and 0.87$\,$mm (bottom), respectively. The left and middle panels show contours identical to that in Figure \ref{observation}. The right panels show residual contours at $(-35,-5,5,35)\times\sigma_{\rm I}$}
    \label{obs_model}
\end{figure*}

\subsection{Disk Model}
\label{diskmodel}
In this section, we construct an axisymmetric disk model that best reproduces Stokes $I$ images of the polarization observations at 3.1$\,$mm, 1.3$\,$mm, and 0.87$\,$mm, which corresponds to ALMA observation band 3, 6, and 7, respectively. 

First, we assume a single radiation source to be the protostar at the disk center. The protostar has the effective temperature $T_{\star}$ = 4000 K and radius $R_{\star}=7R_\odot$, which corresponds to a luminosity $L_{\star}\sim$$11L_{\odot}$ \citep{1999ApJ...519..257M, 2016ApJ...816...25P, 2017A&A...607A..74L}. The radiation field and dust temperature within the disk are then calculated using radiative transfer calculation. Since temperature is a strong function of grain size, we fix $a_{\rm max}$ = 100 ${\mu}$m for temperature calculation to fix the disk's temperature profile. 

Next, we build the surface density profile for the disk. It is commonly estimated by using the spectral energy distribution (SED) fitting. However, it is only simple when the temperature profile is fixed or can be analytically approximated. In our study, the main source of the disk temperature is radiative heating requiring radiative transfer calculation which is computationally costly. Thus, for simplicity, we assume a surface density profile taken from \cite{2011ApJ...741....3K},
\begin{align}
    \Sigma \propto \left(\frac{R}{R_c}\right)^{-\gamma} \text{exp}\left[\left(\frac{R}{R_{\text{c}}}\right)^{2-\gamma}\right]\text{,}
\end{align} 
where $R$ is the distance to the disk center, $R_{\rm c}$ = 78.9 AU and $\gamma = -0.2$. Note that the dust surface density profile is assumed to have a smooth radial distribution and ignore all the gaps and rings observed with higher spatial resolution in \cite{2015ApJ...808L...3A}; this is the case for the lower resolution of ALMA polarimetric observations that we are attempting to reproduce.

The vertical dust density is assumed to follow Gaussian distribution, 
\begin{align}
    \rho = \frac{\Sigma}{\sqrt{2\pi}h_{\rm d}} \text{exp}\left(\frac{-z^2}{2h_{\rm d}^2}\right),
\end{align}
where $z$ is the height from the disk midplane and $h_{\rm d}$ is the disk dust scale height whose profile is taken as 
\begin{align}
    h_{\rm d}(R) = h_{0}\left(\frac{R}{100\;\text{AU}}\right)^{\beta},
\end{align}
with $h_0$ = 10 AU being the scale height at 100$\,$AU from the center of the disk, and $\beta$ = 1.15 the exponent factor characterizing the flaring effect of the disk (\citealp{2016ApJ...816...25P, 2017A&A...607A..74L}). However, HL Tau was studied to have a high degree of dust settling based on the appearance of rings and gaps on millimeter observations \citep{2016ApJ...816...25P}. Dust settling is expected to have a significant effect on polarization due to self-scattering \citep{2017MNRAS.472..373Y}. Hence, in the calculation for self-scattering only, we take into account the effect of dust-settling by dividing the scale height $h_{\rm d}$ by a factor \mbox{$f_{\rm settle}=10$}, which is defined as the dust-settling factor from the disk scale height, following \cite{2016ApJ...820...54K}. In this calculation, the temperature distribution is taken identical to that from the radiative heating calculation with the original scale height. We assume that the disk geometrical thickness would have a minor effect on the intrinsic dust thermal emission at (sub)millimeter wavelength for simplicity.

Assuming the fiducial grain size of $a_{\rm max}$ = 100 ${\mu}$m, we take the total dust mass of the system $M_{\rm d}$ as a calibrating parameter to reproduce the three Stokes $I$ emissions at the three wavelengths mentioned earlier. $M_{\rm d}$ = $1.3\times10^{-2}M_{\odot}$ is found to be the best-fitted value, which produces total emissions comparable to the observations as shown in Figure \ref{obs_model}. This value is similar to the level of $M_{\rm d} \sim 1\times10^{-2}M_{\odot}$ as estimated in \citet{2021ApJ...908..153M} and \citet{2023ApJ...953...96Z}, in the case where grain size at $\sim100\,\mu$m is employed. Although the residuals are still significant, we stress that our aim is to reproduce axisymmetric continuum emissions for the whole three wavelengths with a simple model and ignore the asymmetries of the disk in the later modeling. Therefore, we will not further search for more optimal parameter sets. For the gas part, under the common assumption of gas-to-dust ratio equal to 100 (i.e., the typical value in the ISM), the total disk mass (gas+dust) would be $M_{\rm disk}\,\sim\,1.3\,M_{\odot}$. This total disk mass is similar to that of the central star ($M_{\star}\,=\,1.7\,M_{\odot}$, \citealp{2016ApJ...816...25P}), which would imply that the disk is extremely unstable and therefore, unrealistic. On the other hand, \citet{2011ApJ...741....3K}, \citet{2017A&A...607A..74L}, and \citet{2019ApJ...883...71C} give the value of total dust mass is $M_{\rm d}\,=\,1\times10^{-3}M_{\odot}$. Assuming gas-to-dust ratio equal to 100, the value of total disk mass is $M_{\rm disk}\,=\,0.1\,M_{\odot}$, which is an order-of-magnitude lower than our estimated value and much more reasonable. Thus, to obtain the gas density, we normalize the density profile with $M_{\rm g}\,=\,0.1\,M_{\odot}$, which is equivalent to a gas-to-dust ratio equal to $\sim\,10$.

The radiation field and temperature profile resulting from radiative transfer calculation with this disk model are shown in Appendix \ref{radfield_app}. We assume gas temperature to be the same as the dust temperature for simplicity.

Concerning the magnetic field, we assume that it is dominated by toroidal component as indicated in \cite{2015A&A...574A..68F} with the strength's profile as estimated in \cite{2011ApJ...739...50B} assuming the accretion process within the disk is solely driven by magnetorotational instability in the active layer.
\begin{align}
    B = 1.0\;\text{G} \times \left(\frac{\dot{M}}{10^{-8}\;M_\odot\;\text{yr}^{-1}}\right)^{-1/2}\left(\frac{R}{1\; \text{AU}}\right)^{-11/8}
\end{align}
where $\dot{M} = 1.6 \times 10^{-7} M_\odot\;\text{yr}^{-1}$ is the accretion rate measured by \cite{2004ApJ...616..998W}. 

\begin{figure*}
    \centering
    \includegraphics[scale=0.5]{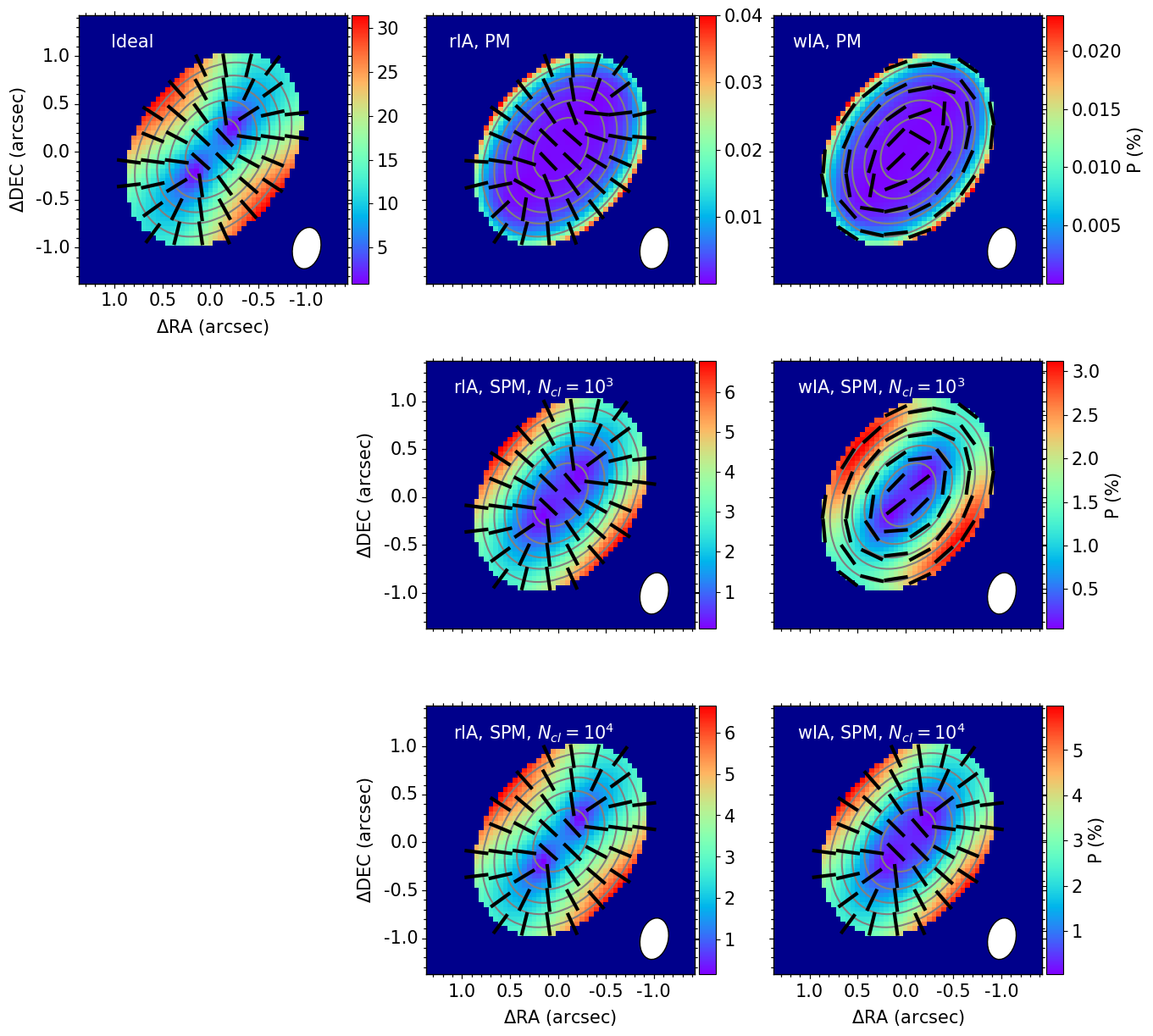}
    \caption{Top left image represents the polarization map of the ideal scenario where all of the grains achieve perfect internal and external alignment (i.e., $f_{\rm high-J}\,=\,1$). The central and right panels include the effects of slow internal relaxation and enhanced (super)paramagnetic relaxation. Central panels show the case of right internal alignment (rIA), while the right ones correspond to wrong internal alignment (wIA). Three different dust models are taken into consideration with paramagnetic grains (top) and superparamagnetic grains which have $N_{\rm cl}=10^3$ (middle) and $10^4$ (bottom), respectively.}
    \label{align_result}
\end{figure*}

\section{Numerical Results}
\label{sec:results}
\subsection{Effect of Grain Magnetic Properties on Disk Polarization}
\label{ncl_pol_result}
Here, we investigate the effect of grain magnetic properties on disk polarization due to the MRAT mechanism at $\lambda$ = 3.1$\,$mm with fiducial grain size \mbox{$a_{\rm max}=100\;\mu$m}. Note that although the same disk and dust models as described in Section \ref{model} are applied, we temporarily ignore the effect of self-scattering for this study.

In Figure \ref{align_result}, we show the resulting polarization degree maps overlaid with polarization vectors considering three different scenarios of internal alignment. The different grain alignment scenarios considered in this section are summarised in Table \ref{appen_alig}.

Initially, we examine the ideal scenario (top left panel) in which all grains have perfect internal alignment (i.e., the fraction of grains aligned at high-$J$ attractors is \mbox{$f_{\rm high-J}\,=\,1$}). In this case, grains are expected to have their longer axis aligned perpendicular to the toroidal magnetic field with optimal efficiency, which results in the radial polarization pattern with a very high polarization degree of up to the level of $\sim$10\%.

In the subsequent scenarios, we take into account the effect of slow internal relaxation and enhanced (super)paramagnetic alignment by MRAT. We consider both cases of right and wrong internal alignment (middle and right panels). The effect of paramagnetic grains (top panels) and superparamagnetic grains with $N_{\rm cl}\;=\;10^3$ and $10^4$ (middle and bottom panels) are also included for comparison. 

In the case of right internal alignment (middle panels), the polarization vectors are the same as in the ideal case, yet the polarization degree is significantly lower. When grains are made of paramagnetic material, the polarization degree is $\sim$0.01\%. In this case, the alignment of very large grains is extremely weak and easily dwarfed by other mechanisms such as $k$-RAT \citep{2017ApJ...839...56T}. Hence, it would be impossible to observe such a disk polarization in reality. For superparamagnetic grains, the alignment is considerably more efficient (i.e., the polarization degree is two orders of magnitude higher than in the paramagnetic grains scenario). With increasing $N_{\rm cl}$, higher values of $f_{\rm high-J}$ (i.e., more efficient grain alignment) are expected due to increasing magnetic relaxation \citep{2016ApJ...831..159H}. However, when grains are aligned with slow internal relaxation in the same direction, their internal alignment efficiencies are the same whether at high-$J$ or low-$J$ attractors. Thus, the polarization degrees observed at $N_{\rm cl}\,=\,10^3\;\text{and}\;10^4$ are seemingly identical.

In the case of wrong internal alignment (right panels), the polarization morphology is more diverse. Grain alignment is weak for paramagnetic grains with polarization degrees as low as $\sim$0.01\%. The polarization vectors show an azimuthal pattern owning to the inefficient internal alignment of grains' $\boldsymbol{\hat{a_1}}$ perpendicular to $\boldsymbol{J}$ at low-$J$ attractors with slow internal relaxation and external alignment of $\boldsymbol{J}$ along $\boldsymbol{B}$. superparamagnetic grains, conversely, induce a higher polarization degree with increasing $N_\text{cl}$. However, the polarization pattern alters when we keep increasing $N_\text{cl}$. Indeed, the increased level of iron inclusion also causes the increased abundance of grains at high-$J$ attractors (i.e., higher $f_\text{high-J}$) and more efficient Barnett relaxation, which would drive them to have more efficient internal alignment with the grains' longer axis being aligned perpendicular to the magnetic field. As a result, the polarization vectors would eventually have the radial direction when $N_{\rm cl}\;=\;10^4$.

\begin{table*}
\centering
\caption{Summary of grain models considered in Section \ref{ncl_pol_result}.}
\label{appen_alig}
\begin{tblr}{
colspec={|>{\centering\arraybackslash}m{1.4cm}|c|c c c|}
}
    \hline
    \SetCell[c=2,r=2]{}  && \SetCell[c=3]{} Grain Alignment Model && \\
    \hline
    &  & perfect IA & right IA & wrong IA \\
    \hline
    \SetCell[r=4]{} Grain Magnetic Property  & $\chi\,\sim\,\infty$ & Ideal & \SetCell[c=2]{} &  \\
    & PM grains & \SetCell[r=3]{} & rIA, PM & wIA, PM\\
    & SPM grains ($N_{\rm cl}\,=\,10^3$) &  & rIA, SPM, $N_{\rm cl}\,=\,10^3$ & wIA, SPM, $N_{\rm cl}\,=\,10^3$\\
    & SPM grains ($N_{\rm cl}\,=\,10^4$) &  & rIA, SPM, $N_{\rm cl}\,=\,10^4$ & wIA, SPM, $N_{\rm cl}\,=\,10^4$\\
    \hline
\end{tblr}
\begin{minipage}{14cm}
\vspace{0.3cm}
\small  \textbf{Notes}. $\chi$ is the grain magnetic susceptibility, which takes a very large value for the ideal case. Grain alignment models right IA/wrong IA denotes the models where grains are aligned with right internal alignment/wrong internal alignment when they are at low-$J$ attractors and slow internal relaxation.
\end{minipage}
\vspace{0.5cm}
\end{table*}

\begin{table*}
\centering
\caption{Summary of the correlation between the Models and Observations.}
\label{MOP}
\begin{tblr}{
colspec={|>{\centering\arraybackslash}m{1.2cm}|c c|c c|c c|c c|}
}
    \hline
    \SetCell[c=3,r=2]{} &&& \SetCell[c=6]{} \textbf{Grain Magnetic Property (SPM)} && \\
    \hline
     &&& \SetCell[c=2]{}$N_{\rm cl}\,=\,10^2$ && \SetCell[c=2]{} $N_{\rm cl}\,=\,9\,\times\,10^2$ && \SetCell[c=2]{} $N_{\rm cl}\,=\,10^4$ &\\
    \hline
    \SetCell[r=9]{} \textbf{Grain Size}  & \SetCell[r=3]{} $\lambda\,=\,3.1\,$mm & $a_{\rm max}\,=\,50\,\mu$m & 0.91 & 0.24 & 0.85 & 0.61 & $-0.77$ & 1.77 \\
    && $a_{\rm max}\,=\,90\,\mu$m & 0.61 & 0.03 & \textbf{0.93} & \textbf{0.69} & $-0.76$ & 2.18 \\
    && $a_{\rm max}\,=\,200\,\mu$m & 0.07 & $-0.26$ & 0.39 & $-0.34$ & $-0.52$ &  1.83\\
    \hline
    & \SetCell[r=3]{} $\lambda\,=\,1.3\,$mm & $a_{\rm max}\,=\,50\,\mu$m & 0.91 & 0.24 & 0.86 & 0.25 & $-0.22$ & $-0.44$ \\
    && $a_{\rm max}\,=\,80\,\mu$m & 0.83 & 0.38 & \textbf{0.93} & \textbf{0.63} & 0.17 & $-0.82$ \\
    && $a_{\rm max}\,=\,100\,\mu$m & 0.84 & 0.52 & 0.87 & 0.64 & 0.50 & $-0.38$ \\
    \hline
    & \SetCell[r=3]{} $\lambda\,=\,0.87\,$mm & $a_{\rm max}\,=\,50\,\mu$m & 0.97 & 0.04 & 0.93 & 0.06 & 0.58 & $-0.87$ \\
    && $a_{\rm max}\,=\,60\,\mu$m & 0.97 & 0.06 & \textbf{0.93} & \textbf{0.09} & 0.77 & $-0.57$\\
    && $a_{\rm max}\,=\,100\,\mu$m & 0.88 & 0.50 & 0.82 & 0.30 & 0.93 & 0.85 \\
    \hline
    \SetCell[c=3]{} &&& $\overline{AM}$ & $c_{\rm P}$ & $\overline{AM}$ & $c_{\rm P}$ & $\overline{AM}$ & $c_{\rm P}$ \\
    \hline
\end{tblr}

\begin{minipage}{15cm}
\vspace{0.3cm}
\small  \textbf{Notes}. We only consider superparamagnetic (SPM) grains in the model. Each column represents the model with a different value of the number of iron atoms per cluster ($N_{\rm cl}$). Each row represents the model with different grain sizes at each wavelength. Each model with a couple value of $N_{\rm cl}$ and $a_{\rm max}$ has two validating factors $\overline{AM}$ and $c_{\rm P}$. $\overline{AM}$ is the average value of alignment measure across the image. $c_{\rm P}$ is the correlation factor of pixel-by-pixel polarization degree between the models and observations. The closer the values of $\overline{AM}$ and $c_{\rm P}$ to unity, the better the model compared to the observations.
\end{minipage}
\vspace{0.5cm}
\end{table*}

\begin{figure*}[!ht]
    \centering
    \includegraphics[scale=0.47]{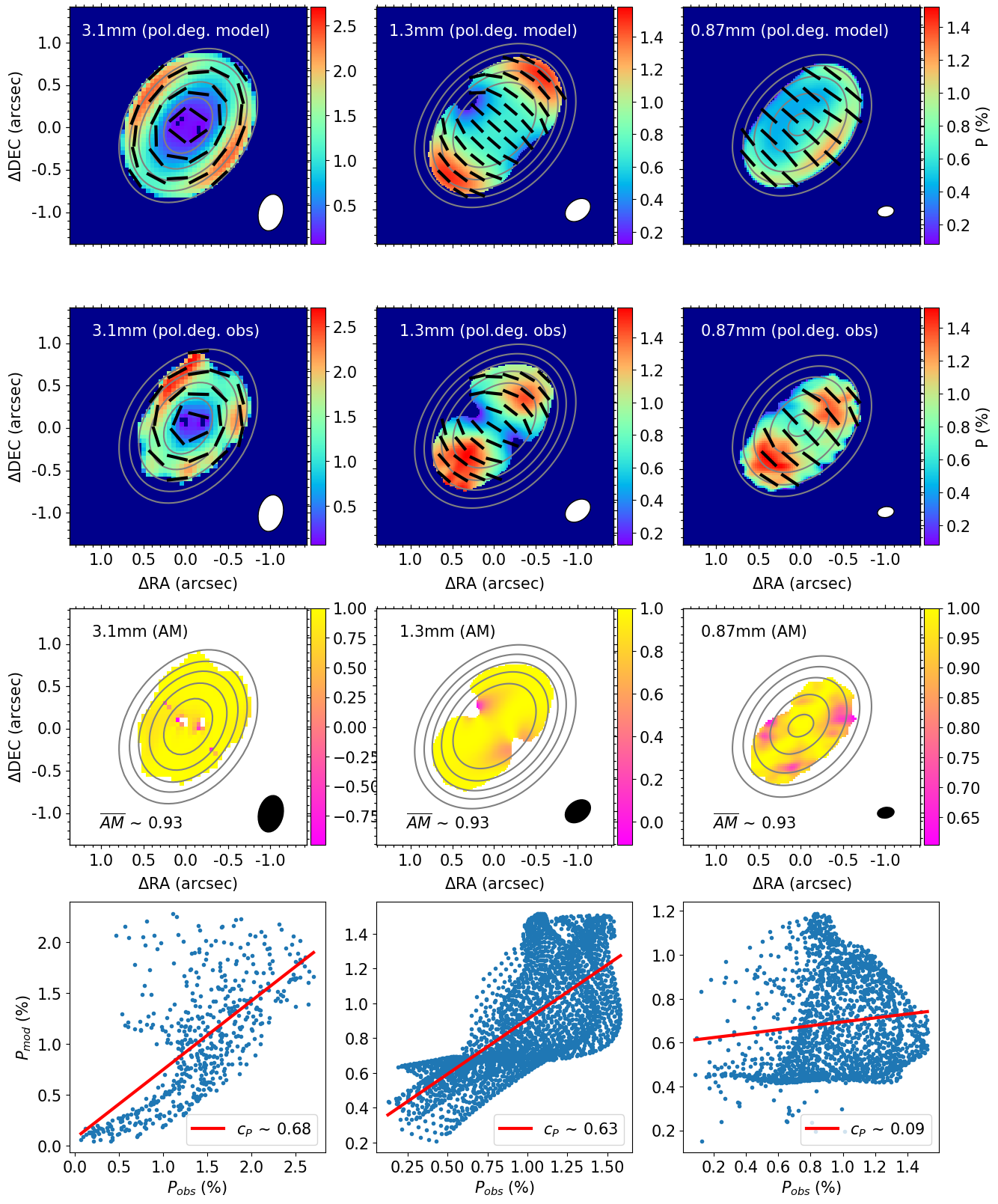}
    \caption{The top panels show the total dust polarization induced by both self-scattering and aligned grains via the MRAT mechanism. These results are obtained for grains aligned with wrong internal alignment, $N_{\rm cl} \sim 9\times10^2$; and $a_{\rm max}$ = 90, 80, and 60 $\mu$m at $\lambda$ = 3.1, 1.3, and 0.87$\,$mm, respectively. Second-row panels show the ALMA polarization observations at the corresponding wavelengths. The third-row panels show the maps of alignment measure ($AM$), which is an indicator of the similarities between the polarization pattern from the model and that from the observations. The bottom panels are the pixel-to-pixel correlations between the polarization degree from the model and that from the observations. The blue dots correspond to the polarization degree data points of each pixel. The red lines are linear fits to the correlations. The values of ($\overline{AM},\,c_{\rm P}$) of the model are $\sim$ (0.93, 0.68), (0.93, 0.63), and (0.93, 0.09) at $\lambda$ = 3.1, 1.3, and 0.87$\,$mm, respectively.}
    \label{final_result}
\end{figure*}

Note that the features of disk polarization only induced by grain alignment observed above also hold for observations at both wavelengths $\lambda = 1.3$ and 0.87$\,$mm.

\subsection{Mixture of MRAT Grain Alignment and Self-scattering: Final Results}  
\label{resultf}

\cite{2021ApJ...908..153M} concluded that a superposition of azimuthally aligned grains and self-scattering can produce the HL Tau disk polarization. In the previous Section \ref{ncl_pol_result}, we demonstrate that only the grain alignment model of wrong internal alignment with superparamagnetic grains having $N_{\rm cl}$ at the level of $\sim$$10^3$ can produce the azimuthal polarization morphology and the polarization degree at $\sim$1\%, which are ones of the notable features of the observations and in line with the description from the literature. Hence, in this section, we limit our grain alignment model to this scenario and add the effect of self-scattering to reproduce the observational data.

To validate the agreement between models and observations, we compare separately their polarization degree and polarization patterns. We check the pixel-by-pixel correlation for the polarization degree ($c_{\rm P}$) while computing the Alignment Measure ($AM$) as introduced in \citet{2017ApJ...835...41G} to compare the polarization orientation.
\begin{align}
    AM = 2\left(\text{cos}^{2}\theta_\text{r}-\frac{1}{2}\right),
\end{align}
where $\theta_\text{r}$ = $|\theta_\text{mod}\;-\;\theta_\text{obs}|$ is the angle difference between the model and observation. The scale of $AM$ is from $-1$ to 1, where $AM=1$ corresponds to a perfect match of polarization vectors and $AM=-1$ indicates that the two vectors are perpendicular to each other. We then take its average value across the image $\overline{AM}$ to represent the overall correlation between polarization vectors of the model and observations. Finally, the values of $c_{\rm P}$ and $\overline{AM}$ are compared between each model to find the best model for the observations. Note that the closer the values of $c_{\rm P}$ and $\overline{AM}$ to unity, the better the agreement between the model and the observations. Table \ref{MOP} summarizes the values of $c_{\rm P}$ and $\overline{AM}$ of the typical models that we considered.  

As shown in Table \ref{MOP}, the model that best reproduces the disk polarization observed by ALMA is found at $N_{\rm cl}=9\times 10^2$; $a_{\rm max}=$ 90, 80, and 60 $\mu$m for $\lambda$ = 3.1, 1.3, and 0.87$\,$mm, respectively. The disk polarization corresponding to this model is displayed in Figure \ref{final_result}. As expected, very large grains with wrong internal alignment produce azimuthal polarization vectors with strong polarization degrees along the major axis while self-scattering produces both polarization vectors and strong polarization degrees along the disk minor axis. Furthermore, as the probing wavelength decreases, the optical depth of the disk increases, which induces a weaker polarization signature from grain alignment due to dichroic extinction \citep{2022MNRAS.512.3922L}. As a result of their superposition at different wavelengths, the polarization shows a transition from azimuthal to parallel to the disk minor axis patterns as in the observations. 

\section{Discussions}
\label{sec:discussions}

\subsection{Comparison to Previous Studies and Implications}

Since ALMA's first polarimetric detection of HL Tau \citep{2014Natur.514..597S}, there have been many observational and theoretical studies aiming to understand the diversity of dust polarization observed toward the disk. Using the RAT alignment theory, \cite{2017ApJ...839...56T} first performed a theoretical study of grain alignment in protoplanetary disks and concluded that magnetic alignment is impossible for mm-sized grains in this environment, even with a high degree of iron inclusion, consistent with predictions in \citet{2016ApJ...831..159H}. The authors suggested that such large grains can be aligned via radiation direction (i.e., $k$-RAT, \citealt{2007MNRAS.378..910L}), which can produce the observed azimuthal polarization pattern. However, \cite{2017ApJ...839...56T} assumed the right internal alignment for very large grains with slow internal relaxation, but their wrong internal alignment cannot be ruled out \citep{2022AJ....164..248H}. In this paper, our detailed modeling for HL Tau using the updated POLARIS shows that the disk polarization can be reproduced with magnetic alignment via the MRAT mechanism, but very large grains with slow internal relaxation must have wrong internal alignment. Therefore, dust polarization can still trace the magnetic field within protoplanetary disks. Moreover, due to the wrong internal alignment, the polarization vectors are along the magnetic field direction, so that one does not need to rotate the polarization vectors by $90^{\circ}$ to infer the magnetic field, as in the case of right internal alignment.  

However, \cite{2019MNRAS.483.2371Y} and \cite{2021ApJ...908..153M} modeled the HL Tau polarization observed at ALMA Band 3 ($\lambda$ = 3.1$\,$mm) and concluded that $k$-RAT could not explain its polarization pattern, which turns out to be more delicate being elliptical, namely the polarization radiation is required to be emitted only in the disk plane. Therefore, the authors suggest that the HL Tau polarization could be reproduced by effective prolate grains that are aligned with their major axis along the azimuthal direction. However, the reason why the grains could be aligned in such a way is not discussed. In our detailed modeling, such a scenario is the outcome of grain alignment physics in which the grains' longer axes are aligned along their angular momenta due to wrong internal alignment for grains with slow internal relaxation, and the grains' angular momenta are aligned along the direction of the toroidal magnetic field as a result of enhanced superparamagnetic relaxation and Larmor precession for grains with embedded iron inclusions.

\citet{2020A&A...634L..15G} proposed that in the Mie regime ($\rm x\,=\,2\pi a/\lambda\,\sim\,1$), the polarized emission can turn into a negative value, which results in the polarization vectors parallel to the toroidal magnetic field for elongated spheroidal grains aligned with right internal alignment. This idea can provide an alternative to our model in explaining the azimuthal polarization pattern observed in the disk. However, the persistence of the azimuthal polarization pattern across a large range of wavelength, from $\lambda\,=\,0.87\,$mm up to $7\,$mm \citep{2023MNRAS.520.1210L}, means that they must all be well within the Mie regime, or the effective grain size must be up to $a_{max}\,\sim\,1\,$mm. This is in contradiction with our constrained grain size where they are only placed at $\sim\,100\,\mu$m-scaled. Hence, the grain size constraint would have implications in determining whether the ALMA observations are in the Rayleigh or Mie regime. Alternatively, it is also worth noticing that as soon as $\rm \lambda\,>\,2\pi a$, the polarized emission of aligned grains is predicted to become positive. Therefore, observing whether there is a polarization flip to a radial pattern at a shorter wavelength than 0.87$\,$mm or longer than 7$\,$mm can help further clarify which is the correct explanation for the observed disk polarization.

In addition to the polarization properties, grain growth in HL Tau has also been in the spotlight due to the conflict of its constraint from polarization degree and spectral index of thermal emission \citep{2021ApJ...908..153M, 2021ApJ...913..117U, 2023MNRAS.520.1210L, 2023ApJ...953...96Z}. Earlier in the current study, $a_{\rm max}$ was constrained to be different when probing at different wavelengths (90$\,\mu$m, 80$\,\mu$m, and 60$\,\mu$m at $\lambda\,=\,\text{3.1}\,$mm, 1.3 mm, and 0.87$\,$mm, respectively). However, they are all at $\sim\,100\,\mu$m, which is still greatly differs from the 1 mm grain size constrained from spectral index SED fitting. Further discussions are presented in Section \ref{grainsizepop}.

\subsection{Constraints on Iron Abundance within Grains in the form of Clusters}
\label{sphisp}
Our results in Section \ref{sec:results} show that paramagnetic grains produce a very low polarization degree, $\sim$0.01\%, which is much lower than the observed polarization level. On the contrary, grains with embedded iron clusters can produce the observed polarization level. Therefore, iron atoms are unlikely distributed diffusely inside dust grains, but must be present in the form of iron clusters.

In Section \ref{sec:results}, for convenience, we fixed the value of \mbox{$\phi_{\rm sp}$ = 0.1} and varied $N_\text{cl}$ to obtain the different magnetic susceptibility of grains (see e.g. Equation 3 in \citealt{2023MNRAS.520.3788G}). In this section, we treat $\phi_{\rm sp}$ as a free parameter to investigate its effect on disk polarization. For simplicity, we test the effect using $q'_{100}$, which is the polarization degree of Stokes $Q$ in the principal frame at the radius of 100 AU along the disk minor axis. A principal frame is a frame whose ordinate coincides with the disk minor axis. In this way, $q'_{100}$ would have a positive value when the polarization vector is radial and negative when it is azimuthal. We demonstrate the effect at $\lambda$ = 3.1$\,$mm, where alignment is expected to be the dominant component over self-scattering. The value of $-q'_{100}=1.95$ due to the alignment component as decomposed in \cite{2022MNRAS.512.3922L} will be chosen as the reference value. An error of 0.5$\%$ is assigned to represent the uncertainty of the method. 

\begin{figure}
    \centering
    \includegraphics[scale=0.65]{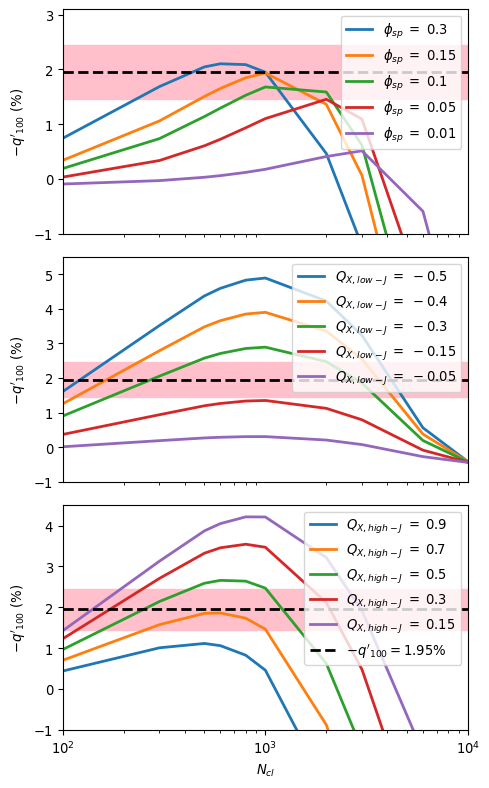}
    \caption{Dependence of polarization degree $-q'$ of Stokes $Q$ due to grain-alignment-only in the principal frame at \mbox{100 AU} along the disk's minor axis on $N_{\rm cl}$. The dashed line corresponds to the value measured in \cite{2022MNRAS.512.3922L} using a plane-slab model with an image of HL Tau at Band 3. The filled area corresponds to an uncertainty of 0.5\%. The top panel takes into consideration different values of $\phi_{\rm sp}$ with the original dust model set-up. The middle panel considers the varying values of $Q_{\rm X,high-J}$ with the extreme scenario when $\phi_{\rm sp}=0.3$ and $Q_{\rm X,low-J} = -0.5$. The bottom panel is plotted with different values of $Q_{\rm X,low-J}$ when $\phi_{\rm sp}$ = 0.3 and $Q_{\rm X,high-J}$ = 0.01. Only the value of $\phi_{\rm sp} \gtrsim$ 0.05; $|Q_{\rm X,high-J}| \lesssim$ 0.8, and $|Q_{\rm X,low-J}| \gtrsim$ 0.15 can reach the threshold.}
    \label{pqx}
\end{figure}

The demonstration is shown in the top panel of Figure \ref{pqx}. Generally, at each value of $\phi_\text{sp}$, the curve describing the dependence of $-q'_{100}$ on $N_\text{cl}$ follows a parabolic shape. The reason behind this is as described in Section \ref{ncl_pol_result} in which increased $N_\text{cl}$ not only increases polarization degree through enhancing external alignment but also increases $f_\text{high-J}$ which drives polarization vector from azimuthal to radial direction. The value of $-q'_{100}$ is at its peak when the effect of superparamagnetic alignment from $N_\text{cl}$ is not negligible so that grains can efficiently align with the magnetic field, yet not too significant so that $f_\text{high-J}$ is small enough to keep the polarization vectors at the azimuthal direction. With the higher value of $\phi_\text{sp}$, this peak shifts to the position corresponding to the lower value of $N_\text{cl}$. 

This test puts a constraint on $\phi_{\rm sp}$ with its minimum value at $\sim$0.05, which corresponds to $\sim$16\% of iron locked inside grains in the form of cluster. Finally, it is crucial to point out that not only our constrained value of $N_{\rm cl}\sim 9 \times 10^2$ is consistent with the $-q'_{100}$ reference when considering $\phi_{\rm sp}$ = 0.1 as in our dust model, but the range of $N_{\rm cl}$ is also found to be at $\sim$$2 \times 10^2$ - $2 \times 10^3$ within the valid values of $\phi_{\rm sp}$.

\subsection{Constraints on Internal Alignment Efficiency for Slow Internal Relaxation}
\label{sINR}

Slow internal relaxation is generally responsible for internal alignment of large dust grains within protoplanetary disks since large grains above $\sim$10 $\mu$m and high density are ubiquitous in these environments \citep{2022ApJ...928..102H}. However, the efficiency of internal alignment for the case of slow internal relaxation is not yet available from theory \citep{2009ApJ...697.1316H, 2022AJ....164..248H}. Usually, the efficiency of internal alignment by slow internal relaxation is represented by two quantities $Q_{\rm X,low-J}$ and $Q_{\rm X,high-J}$. In Section \ref{sec:results}, our modeling results showed that $Q_{\rm X,low-J}$ must have a negative value (i.e., wrong internal alignment) to reproduce the azimuthal polarization pattern observed toward the HL Tau disk. On the contrary, grains at high-$J$ attractors always have the right internal alignment or $Q_{\rm X,high-J}$ is always positive \citep{2009ApJ...697.1316H}. It indicates that the directions of internal alignment of grains aligned at low-$J$ and high-$J$ attractors are opposite. 

Here, we will attempt to empirically constrain the magnitude of $Q_{\rm X,low-J}$ and $Q_{\rm X,high-J}$ for slow internal relaxation using the observed polarization data. To do so, we consider the extreme scenarios where parameters involved in the internal alignment process are set to their respective extremes. We initially fix $\phi_{\rm sp}=\phi_\text{sp,max}=0.3$. To obtain the maximum value of $|Q_{\rm X,high-J}|$, we set $|Q_{\rm X,low-J}|=|Q_{\rm X,low-J,max}|=0.5$. On the other hand, $|Q_{\rm X,high-J}|=|Q_{\rm X,high-J, min}|\sim$ 0.01 is set to constrain the minimum value of $|Q_{\rm X,low-J}|$. Following Section \ref{sphisp}, we use $-q'_{100}$ at $\lambda=3.1$ mm as the referenced polarization degree for the disk polarization. Its value is calculated for different values of $Q_{\rm X,high-J}$ or $Q_{\rm X,low-J}$ and over the range $N_{\rm cl}$ = [$10^2$, $10^4$]. The plots are shown in the middle and bottom panels of Figure \ref{pqx}. By considering the values satisfying the reference value, we can deduce the maximum possible value of $|Q_{\rm X,high-J}|$ is $\sim$0.8, while the minimum value of $|Q_{\rm X,low-J}|$ is at $\sim$0.15.

\subsection{Other Grain Alignment Mechanisms}
Besides MRAT, grain alignment due to interaction of the gas flow and grains, including Gold alignment \citep{1952MNRAS.112..215G} and mechanical torque (MET) alignment with the grain's long axis along the flow velocity (so-called \mbox{$v$-MET}; \citealt{2022AJ....164..248H}) can also produce an azimuthal polarization pattern. 

Gold alignment was thought to be the most promising mechanism to explain the polarization morphology caused by grain alignment in HL Tau \citep{2019MNRAS.483.2371Y}. However, the possibility of the Gold mechanism causing grain alignment in disk environments is highly questionable \citep{2021ApJ...908..153M}. Since the mechanism requires supersonic gas speed to have effectively aligned grains, whereas it is generally subsonic in protoplanetary disks \citep{2007ApJ...669.1085C}, it is unlikely to induce grain alignment within this environment.

$v$-MET, on the other hand, offers a rather promising alternative. If grains with Keplerian motions align with the drift direction with wrong internal alignment, the polarization pattern is also expected to be azimuthal \citep{2022AJ....164..248H}. However, it may not be the case if we take a more thorough inspection. The polarization direction depends on the direction of gas flow in the frame of dust grains, which is a strong function of Stokes number denoting how well dust grains couple to ambient gas. With $a_{\rm max} \sim 100$ $\mu$m, and assuming a gas surface density of $\Sigma_{\rm g}\,\sim$ 5 g cm$^{-2}$, the Stokes number is estimated to be $St \sim 5.4 \times 10^{-3}$. This value is much smaller than unity which indicates that the direction of gas velocity with respect to the grains will be radial (see Figure 2 in \citealp{2019ApJ...874L...6K}). In the case of right internal alignment, a circular polarization pattern is produced. On the other hand, if grains are aligned with the wrong internal alignment, the pattern will be radial. Both cases are not compatible with the elliptical patterns from observations \citep{2019MNRAS.483.2371Y}. Note that the argument still holds even when considering grain size up to 1 mm as constrained in \citet{2019ApJ...883...71C}, which corresponds to $St \sim 5.4 \times 10^{-2} \ll 1$.

Nevertheless, it is crucial to stress that we cannot rule out the possibility of enhanced alignment with grain drift \citep{2022AJ....164..248H}, which is neglected in our model. Since the direction of gas flow with respect to the grains is radial, which coincides with the direction of the radiation beam, the combined effect of grain drift and radiation can significantly improve the efficiency of grain alignment within the disk.

In conclusion, it has been demonstrated that MRAT is the only grain alignment mechanism in the constraint of current understanding on grain alignment theory that can explain the polarization signatures observed in the HL Tau disk by ALMA.

\subsection{Grain Size Population}
\label{grainsizepop}

\begin{figure}[!ht]
    \centering
    \includegraphics[scale=0.5]{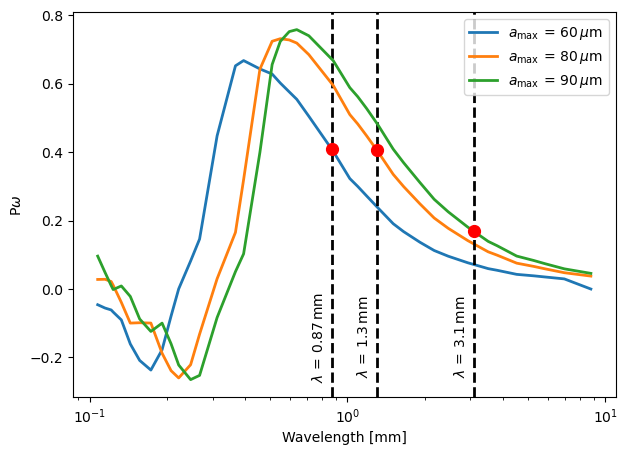}
    \caption{Polarization degree at $90^{\circ}$ scattering times the albedo P$\rm \omega$ against the observing wavelength at the three-grain size models $a_{\rm max}\,=\,60\,\rm \mu m, 80\,\rm \mu m$, and 90$\,\rm \mu$m. The three vertical lines correspond to the three ALMA wavelengths $0.87\,\rm mm, 1.3\,\rm mm$, and $3.1\,$mm, where observations were used as references in our work. The three red dots correspond to the maximum grain size constrained at each observing wavelength.}
    \label{POmega}
\end{figure}

Grain size was constrained primarily from the dust polarization of the self-scattering component. Therefore, in order to address the detailed view of the grain size constraint given in Section \ref{sec:results}, we provide the polarization efficiency from dust scattering of each dust model, as a function of wavelength in Figure \ref{POmega}. Polarization efficiency is represented by the polarization degree at $90^{\circ}$ scattering times the albedo P$\rm \omega$ following \citet{2015ApJ...809...78K}.

We could reproduce the observed dust polarization assuming that the three wavelengths trace different layers of the disk, which contain grains of discrete sizes. In the optically thick region of HL Tau, this is the case because the differential dust-settling effect can cause light at different wavelengths to trace different layers of the disk with longer wavelengths leaning towards the disk midplane where larger grains are expected to be present \citep{2020A&A...640A.122B, 2021ApJ...913..117U}. Indeed, the grain size in our results is constrained at larger values when tracing the dust by longer wavelengths, namely $a_{\rm max}$ = 60, 80, and 90$\,\mu$m at $\lambda$ = 0.87, 1.3, and 3.1$\,$mm, respectively. 

However, this argument is only valid at the optically thick inner region of the disk. The optical depth would decrease with the radial distance from the disk center. Thus, the dust differential settling effect would have a weaker impact on the outer region. Irregular grain structures have been studied to alleviate the grain size conflict between polarization and spectral index SED fitting \citep{2023MNRAS.520.1210L, 2023ApJ...953...96Z}. Moreover, it offers a promising explanation for the spectral dependence of the polarization degree across (sub)millimeter wavelengths as in Figure \ref{POmega}. According to the literature, grain size can be larger than 1$\,$mm depending on the grain porosity. Furthermore, grains with large porosity have less steeper-slope spectral dependence of scattering efficiency, which generates a similar trend for polarization degree resulted from self-scattering. Therefore, it is crucial to incorporate the effect of grain porosity into detailed modeling of grain alignment to have a more complete picture of HL Tau polarization.

\subsection{Disk Polarization with Substructures}

Previously in our original disk model in Section \ref{diskmodel}, the disk is assumed to have a smooth radial density distribution. However, \cite{2015ApJ...808L...3A} has revealed that the disk contains an alternate distribution of rings and gaps, which is a crucial sign for planet formation. \cite{2023Natur.623..705S} are recently able to detect the dust polarization within the substructures at $\lambda\,=\,0.87\,$mm and find the distinguishable features observed in gaps compared to the rings. In the gaps, the authors find the polarization vectors to be azimuthal with a higher polarization degree (up to $\sim$4\%) than within the rings. This is understandable considering within the gaps, the optical depth is at unity, which induces a higher polarization degree due to self-scattering (up to $\sim$3\%, \citealp{2016ApJ...820...54K}). The lower dust opacity also induces lower dichroic extinction, which means a higher polarization degree of the intrinsic emission from aligned grains. The lower gas density in the gaps is also expected to make the grain alignment process more efficient with less effective gas damping \citep{2022AJ....164..248H}. However, too low gas density may also induce more grains aligning with fast internal relaxation and lower the degree of azimuthal polarization. This essentially signifies the importance of having accurate gas density for polarization calculation. Furthermore, with lower gas density, dust and gas are expected to couple more efficiently and the effect of grain alignment due to grain drift would be non-negligible. We attempt to model the disk with substructures in Appendix \ref{appen::gap}. However, it is only a rough estimation and cannot reproduce the entire detailed polarimetric observation as in \cite{2023Natur.623..705S}. One of which is the high polarization degree at the gaps, where their further modeling suggests that the intrinsic polarization could be up to 10-15$\,\%$ without beam smearing effect. This is potentially because the dust vertical settling effect has not been taken into account, which ignores the larger grain size being probed in the gaps due to less opacity. The gas-to-dust ratio contrast in these gaps and rings is also still uncertain \citep{2016ApJ...816...25P, 2019ApJ...880...69Y} and the simple assumption of constant gas-to-dust ratio throughout the disk is no longer acceptable. The accumulation of magnetic field inside the gaps also needs to be considered if they were formed due to magneto-hydrodynamic (MHD) effects \citep{2015A&A...574A..68F}. Moreover, the effect of mechanical torque alignment \citep{2022AJ....164..248H} has not been taken into account. Its incorporation into modeling still faces difficulty due to its significant sensitivity to grain shape \citep{2023A&A...674A..47R}. Hence, additional studies are needed to fully reproduce the high-resolution HL Tau dust polarization observed by ALMA.  

\subsection{Caveats}
Our model is subject to the unresolved grain shape inconsistency. While we calculate the direct polarized thermal emission from aligned elongated dust grains, spherical grains are used in the MCRT calculations. This is due to the difficulties in computing the optical properties of non-spherical dust grains, especially its scattering property.

The assumed geometrically thick disk in intrinsic dust thermal emission calculation may cause unwanted near-far side asymmetry in the disk emission at optically thick region due to more dichroic extinction in the near side of the disk. However, we ignore this effect due to the nature of our simple disk model setup and the preferred need to use a precise approximation of the radiation field and gas density for grain alignment calculation.

\section{Summary}
\label{summary}
The origin of the polarized emission of HL Tau at different wavelengths has been debated in previous studies. In this work, through our detailed multi-wavelength modeling, the following results are found.
\begin{enumerate}
    \item The ideal scenario of magnetic alignment significantly overestimates the polarization degree compared to the realistic cases. In the realistic scenario, especially when grains are aligned at low-$J$ attractors with slow internal relaxation, grains can either align with wrong or right internal alignment, which manifests in azimuthal or radial polarization patterns, respectively. Additionally, paramagnetic grains cannot reproduce the observed polarization data and thus cannot be the primary dust material in HL Tau.
    \item A mixture of self-scattering and MRAT grain alignment mechanism where grains are aligned with wrong internal alignment at low-$J$ attractors and slow internal relaxation can reproduce the dust polarization observed in HL Tau by ALMA at (sub)millimeter wavelengths.
    \item Grains are shown to be made of superparamagnetic materials with the number of iron atoms inside each cluster is $N_{\rm cl} \sim 9\times10^2$. The abundance of embedded iron inside grains identified from our model is at $\sim$16\% minimum.
    \item By assuming grain sizes probed to be $\lambda$-dependent, maximum grain sizes are constrained at $a_{\rm max} \sim$  60, 80, and 90$\,\mu$m when probing at $\lambda$ = 0.87, 1.3, and 3.1$\,$mm, respectively. 
    \item MRAT is found to be the only mechanism in the contemporary grain alignment theory that can explain the polarization morphology caused by the grain alignment component within the HL Tau disk.
\end{enumerate}
\section*{Acknowledgements}
    We thank the anonymous referee for his/her constructive comments that helped improve the presentation of this paper. We thank Ph.D. candidate Nguyen Chau Giang (KASI) for providing the updated POLARIS and significant discussion about the code along with its embedded grain alignment physics. N.T.T. is supported by a grant from the Simons Foundation to IFIRSE, ICISE (916424, N.H.). N.T.P. and P.N.D. are funded by Vingroup Innovation Foundation (VINIF) under project code VINIF.2023.DA.057. N.B.N. was funded by the Master, Ph.D. Scholarship Programme of Vingroup Innovation Foundation (VINIF), code VINIF.2023.TS077. 
    
    This paper makes use of the following ALMA data: ADS/JAO.ALMA$\#$2016.1.00115.S, ADS/JAO.ALMA$\#$2016.1.00162.S,  \par 
    \noindent ADS/JAO.ALMA$\#$2019.1.01051.S. ALMA is a partnership of ESO (representing its member states), NSF (USA), and NINS (Japan), together with NRC (Canada), MOST and ASIAA (Taiwan), and KASI (Republic of Korea), in cooperation with the Republic of Chile. The Joint ALMA Observatory is operated by ESO, AUI/NRAO, and NAOJ.
\software{POLARIS}
\appendix

\section{Radiation Field and Dust Temperature}
\label{radfield_app}
\begin{figure*}[!ht]
    \centering
    \includegraphics[scale=0.5]{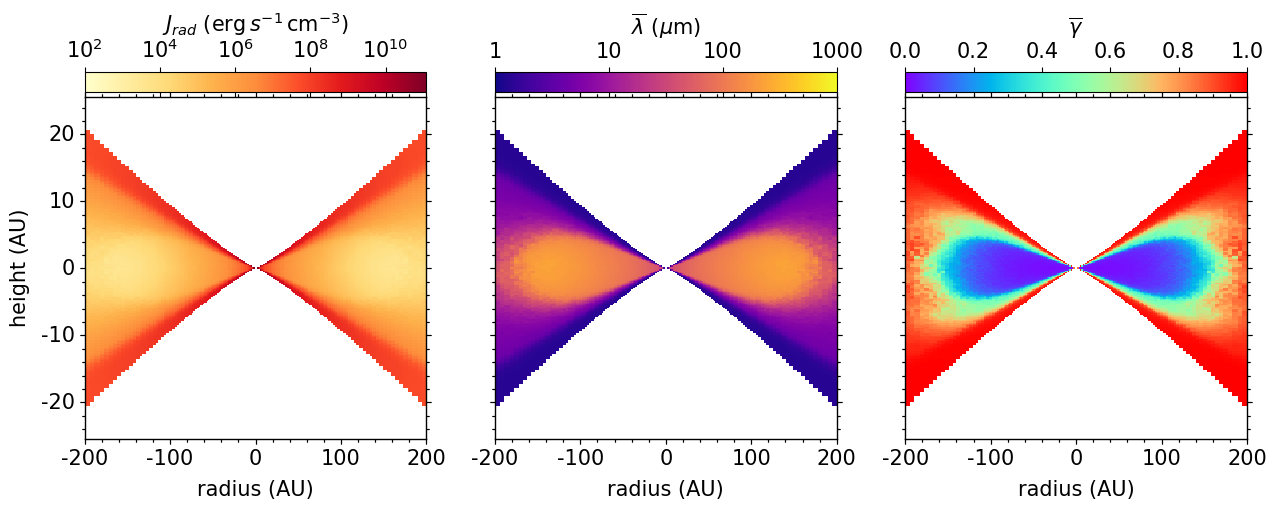}
    \caption{From left to right: Radiation flux $J_\text{rad}$, spectrum weighted wavelength $\overline{\lambda}$, and spectrum weighted anisotropy parameter $\overline{\gamma}$. The calculation is made assuming a single radiation source to be the central protostar with luminosity $L_{\star} \sim 11 L_{\odot}$.}
    \label{rad_field}
\end{figure*}

With the radiation source at the disk center, the radiation flux $J_{\rm rad}$, spectrum weighted wavelength $\overline{\lambda}$, and anisotropy parameter $\overline{\gamma}$ (defined in \citealp{2007ApJ...663.1055B}) are calculated and illustrated in Figure \ref{rad_field}.

While stellar radiation dominates the upper layers of the disk, the inner disk region is dominated by emission from warm dust from the surface and intermediate layers. The upper layers of the disk correspond to the disk photosphere, where it can be observed with IR/NIR observations. The inner disk region, on the other hand, can be traced by radio observations, which is the case in our modeling. Calculations show that the radiation at this region is dominated at $\overline{\lambda}\sim$ 140$\;\mu$m, with anisotropy parameter $\overline{\gamma}\sim 0.1$. Therefore, only about 10$\%$ of the radiation flux primarily at $\overline{\lambda}$ $\sim$ 140 $\mu$m can contribute to RATs towards the grains within the disk because only the anisotropic radiation is important for RAT alignment.

In our model, we also use radiative heating to calculate the temperature profile of the disk. Our radiative transfer calculation yields a temperature profile at the disk mid-plane of

\begin{align}
    T(R) = 347\left(\frac{R}{1\,\text{AU}}\right)^{-0.64}~{\rm K}.
\end{align}

\section{Disk model with ring/gap structures}
\label{appen::gap}
Throughout this work, we adopted the smoothed structure of the HL Tau disk. However, \citet{2023Natur.623..705S} reported ring and gap structures with the new ALMA observations. They found that the polarization degree is much higher in the gaps than in the rings, and the polarization vectors are azimuthal.

This section aims to understand this newly observed phenomenon based on the same physical model. We first mimic the disk structure at $\lambda\,=\,0.87\,$mm as in \citet{2023Natur.623..705S}, which yields the surface density profile
\begin{align}
    \Sigma \propto \Sigma_{a}\left(\frac{R}{R_0}\right)^{-p} \text{exp}\left[-\left(\frac{R}{R_{\text{c}}}\right)^{1.5}\right] + \sum_{i=1}^{N} \Sigma_{i}\,\text{exp}\left[-\frac{1}{2}\left(\frac{R-C_{i}}{W_{i}}\right)^2 \right] \text{,}
\end{align}
where $\Sigma_{a}\,=\,2.3$, $R_{0}\,=\,10\,$AU, $p\,=\,0.5$, and the parameters for the rings are given in Table \ref{table::gap}. In this investigation, instead of using temperature calculated from radiative transfer calculation, for simplicity, we adopt directly the temperature profile as in the literature: $T(R)\,=\,110\,\times\,[R/(10\,AU)]^{-0.5}$. We then normalize the dust density by the total dust mass of $M_{\rm d} \sim 5 \times 10^{-3} M_{\odot}$ to produce a comparable Stokes $I$ emission at $\lambda = 0.87$ mm as in \cite{2023Natur.623..705S}. The gas density, on the other hand, is obtained by normalizing the density profile with the total gas mass of $M_{\rm g}\,=\,0.1\,M_{\odot}$ similar to that done in Section \ref{diskmodel}. This is equivalent to a gas-to-dust ratio of $\sim\,20$.

\begin{table*}[!ht]
    \centering
    \caption{Parameters for the identified rings in HL Tau.}
    \label{table::gap}
    \begin{tabular}{c c c c}
           \hline
           Ring number & $\Sigma_{i}$ & $C_{i}$ [AU] & $W_{i}$ [AU] \\
           \hline
           1 & 8 & 24 & 4 \\
           2 & 5 & 39 & 3 \\
           3 & 5 & 49 & 3 \\
           4 & 8 & 59 & 2 \\
           5 & 3 & 73 & 4 \\
           6 & 8 & 88 & 3 \\
           7 & 8 & 102 & 2 \\
           8 & 8 & 116 & 2 \\
           \hline
    \end{tabular}
\end{table*}

\begin{figure*}[!h]
    \centering
    \includegraphics[scale=0.8]{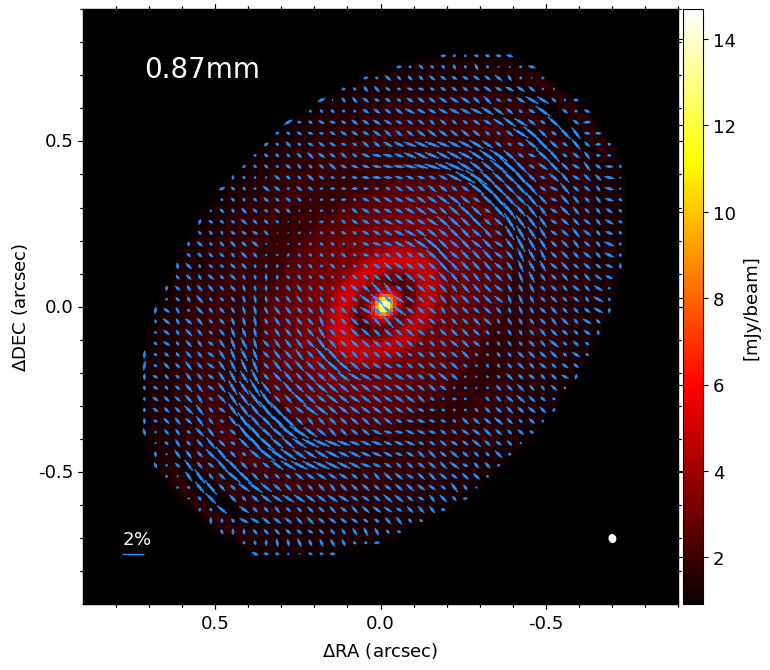}
    \caption{Prediction of the toy model of the HL Tau disk with the ring-gap structures at $\lambda\,=\,0.87\,$mm. The polarization degree is higher in the gap, and the polarization orientation becomes azimuthal.}
    \label{fig::gap}
\end{figure*}

For the polarization calculation, we adopt the same dust models as constrained in Section \ref{resultf} with $N_{\rm cl}=9 \times 10^2$, and $a_{\rm max}=60\,\mu$m. The predicted polarization degree and polarization angles with the ring/gap disk are shown in Figure \ref{fig::gap}. Even with the simple dust model, we can spot differences between the polarization observed in the rings and gaps. In the gaps, we found that the polarization degree is higher, and the polarization caused by the self-scattering component becomes less dominant, resulting in a more azimuthal polarization pattern. These features are consistent with the new observations in HL Tau.

\newpage
\bibliography{ref}{}
\end{document}